\documentclass[aps,pre,amsmath,twocolumn,showpacs]{revtex4}
\usepackage{amsmath}
\usepackage{amsfonts}
\usepackage{graphicx}
\usepackage{hyperref}
\usepackage{mathtools}
\usepackage{verbatim}
\usepackage{epstopdf}
\usepackage{subfigure}
\usepackage{latexsym}
\usepackage{dcolumn}
\usepackage{epsf}
\usepackage{float}
\DeclarePairedDelimiter\abs{\lvert}{\rvert}

\usepackage{color}

\begin{document}
\title{Transition to synchrony in degree-frequency correlated Sakaguchi-Kuramoto model}
\author{Prosenjit Kundu$^1$} 
\author{Pitambar Khanra$^1$}
\author{Chittaranjan Hens$^2$}
\author{Pinaki Pal$^1$}
\email{pinaki.pal@maths.nitdgp.ac.in}
\affiliation{$^1$Department of Mathematics, National Institute of Technology, Durgapur 713209, India}
\affiliation{$^2$Department of Mathematics, Bar-Ilan University, Ramat-Gan 52900, Israel}
\date{\today}

\begin{abstract}
We investigate transition to synchrony in degree-frequency correlated Sakaguchi-Kuramoto (SK) model on complex networks both analytically and numerically. We analytically derive self-consistent equations for group angular velocity and order parameter for the model in the thermodynamic limit. Using the self-consistent equations we investigate transition to synchronization in SK model on uncorrelated scale-free (SF) and Erd\H{o}s-R\'{e}nyi (ER) networks in detail. Depending on the degree distribution exponent ($\gamma$) of SF networks and  phase-frustration parameter, the population undergoes from first order transition (explosive synchronization (ES)) to second order transition and vice versa. In ER networks transition is always second order irrespective of the  phase-lag parameter. We observe that the critical coupling strength for the onset of synchronization is decreased by phase-frustration parameter in case of SF network where as in ER network, the phase-frustration delays the onset of synchronization. Extensive numerical simulations using SF and ER networks are performed to validate the analytical results. An analytical expression of critical coupling strength for the onset of synchronization is also derived from the self consistent equations considering the vanishing order parameter limit.
   
\end{abstract}

 \maketitle
\section{Introduction}
The phenomenon of synchronization in complex systems consisting of a large ensemble of interacting units has attracted  great attention of the researchers due to its ubiquity in natural as well as artificial systems~\cite{Pikovsky,Boccaletti-physrep,Arenas-physrep,Kurths-Physreport}. Examples include flashing of fireflies~\cite{buck:qrb_1988}, applauding persons~\cite{neda:pre61_2000}, systems describing circadian rhythms in animals~\cite{yamaguchi:sc302_2003}, Josephson junction arrays~\cite{wissenfeld:pre_1998}, moving pedestrians on footbridges~\cite{strogatz:nature_2005} and many more~\cite{dorflar:automatica_2014}. Mathematical modeling of these complex systems often involves the important steps of capturing dynamics of the interacting units by nonlinear oscillators and interaction by suitable coupling functions~\cite{boccaletti:pr_2006}.
However, in 1975, Kuramoto~\cite{Kuramoto:1975} showed that in many cases, when the interaction among the units are weak, emergent phenomenon like synchronization in complex systems can successfully be described by an ensemble of coupled phase oscillators~\cite{Kuramoto}. In its simplest setting, the Kuramoto model~\cite{Kuramoto,Bonilla:rmp2005} consists of an ensemble of phase oscillators with heterogeneous natural frequencies often drawn from a unimodal distribution which are globally coupled through the $\it {sine}$ of their phase differences.

In spite of its simplicity, classical Kuramoto model and its generalizations have been successfully applied in numerous systems of scientific and technological interests including  coupled Josephson junctions~\cite{wissenfeld:pre_1998}, decision making in animal groups~\cite{leonard:2012}, semiconductor laser arrays~\cite{kozyreff:2000} etc. mainly to investigate rhythmicity and synchronization (see the review~\cite{dorflar:automatica_2014,Bonilla:rmp2005} and references therein for details). On the other hand, theoretical investigations of the Kuramoto model and its generalizations largely focus on the transition to synchronization in coupled networks~\cite{dorflar:automatica_2014}. The classical Kuramoto model~\cite{Kuramoto} show second order transition to synchronization at a critical coupling strength under trivial all-to-all network topology when natural frequencies are drawn from unimodal distribution. The study on the influence of nontrivial network topology on synchrony gained momentum after the seminal work of Watts and Strogatz~\cite{watts:nature_1998,strogatz:nature_2001,barabasi:science_1999}. 
Network topologies are found to strongly affect the critical coupling strength for the onset of synchronization yet only second order transition to synchrony were reported~\cite{arenas:pr_2008} until the recent work of Jesus et al.~\cite{jesus:prl106_2011}. 

In their work, Jesus et al.~\cite{jesus:prl106_2011} interestingly considered
degree-frequency correlated scale-free (SF) network topology  and  reported a first order  transition to synchrony or so called explosive synchronization (ES) for the first time in Kuramoto paradigm which is characterized by a sharp jump of the order parameter~\cite{Kuramoto} as the system passes from  incoherence to synchronization~\cite{Boccaletti:physrep2016,Leyva:screport2013}. Considering similar degree-frequency correlated SF networks, Peron {et al.}~\cite{Peron:pre86_2012} determined analytical expression of critical coupling for ES using mean field approach~\cite{ichinomia:pre_2004}. 
Subsequently, Coutinho {\it et al.}~\cite{Coutinho:pre87_2013} derived self-consistent coupled equations for group angular velocity  and order parameter of the system which successfully explained the emergence and annihilation of ES island in a scale-free network environment where frequency and degree  are correlated to each other. In a recent work, Pinto {et al.}~\cite{Pinto:2015} performed rigorous mathematical analysis to investigate ES with partial degree-frequency correlation. 
\begin{figure*}
   \includegraphics[width=\textwidth]{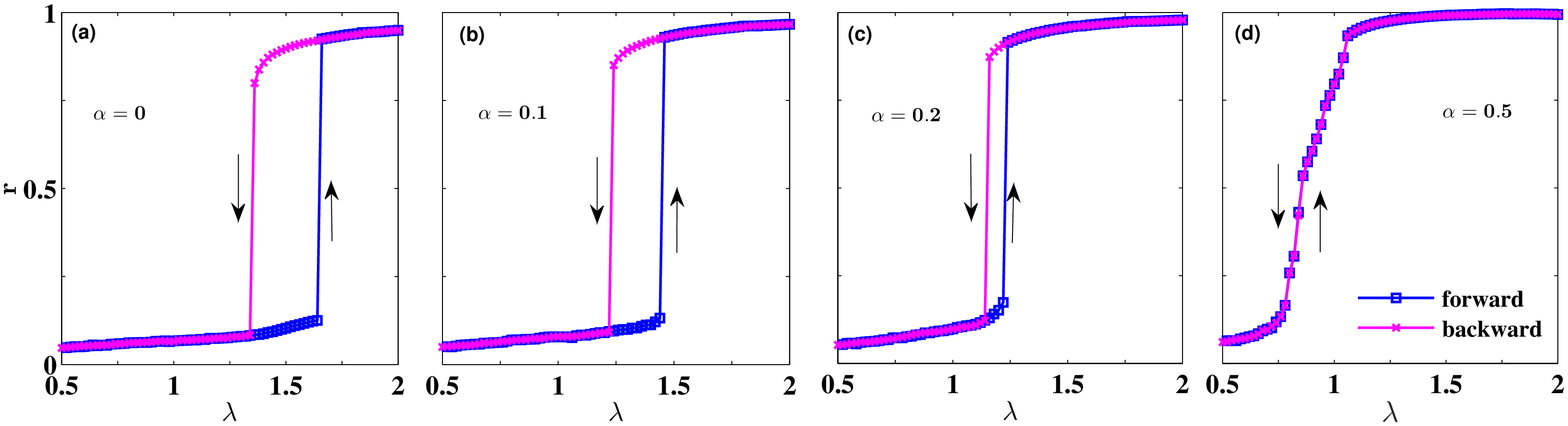}
  \caption{Synchronization diagram of scale-free network of size $N = 1000$, $\gamma = 2.8$ and $\langle k \rangle =30$ for four values of $\alpha$. The lag parameter $\alpha$ apparently inhibiting explosive synchronization.}
  \label{simu:sf_n1000}
\end{figure*}
\begin{figure}[h]
  \includegraphics[height=!,width=8cm]{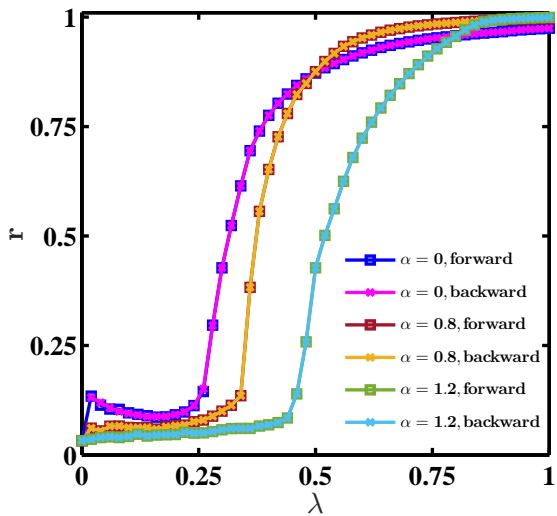}
   \caption{Synchronization diagram of ER network of size $N = 1000$ and $\langle k \rangle =30$ for three values of $\alpha$. Higher phase-lag ( $\alpha$)  {\it delays} the onset of  synchronization.}
  \label{simu:er_n1000}
\end{figure}

The works mentioned in the above paragraph are based on the Kuramoto model on complex networks in absence of phase frustration or phase-lag in the coupling. Kuramoto model in presence of a phase-lag parameter (Sakaguchi-Kuramoto (SK) model~\cite{Sakaguchi:1986}) on complete graph topology shows a traditional second order  transition to synchrony for unimodal frequency distribution ~\cite{Lohe:automatica2015}, while for a non-unimodal frequency distribution it also shows non-universal transition to synchrony~\cite{Omel'chenko:prl2012}. Recently, effect of phase-frustration on transition to synchrony in the degree frequency correlated network has been examined by  Xu {\it et al.}~\cite{Xu:screport2015} in case of simple star-graph motif. However, the whole literature lacks a systematic investigation of desynchronization to synchronization transition in SK model on  degree-frequency correlated complex networks.  

In this paper, we analytically derive self consistent equations involving group angular velocity and order parameter of the degree-frequency correlated SK model on complex networks based on mean field approach using annealed network approximation~\cite{Coutinho:pre87_2013} for investigating transition to synchrony. Coupled analytical expressions of critical coupling strength and group angular velocity at the onset of synchronization are also derived. The analytical theory is demonstrated using scale-free and Erd\H{o}s-R\'{e}nyi networks and compared with numerical simulation results. The qualitative differences in the synchronization transition for  both  networks are explored in detail. We show here that for a large network, the self consistent equations can successfully  describe (i) the emergence and annihilation of explosive synchronization (ES) (from first order to second order transition and vice versa)  and (ii) the transition to synchrony in the parameter space of phase-lag and coupling strength. We show that ES appears only in SF network  when degree distribution exponent $\gamma<3$ whereas  ES never appears in ER network. We also show that the critical coupling strength for the onset of synchronization is decreased by phase-frustration parameter in case of SF network where as in ER network, the phase-frustration delays the onset of synchronization.
 
\section{Sakaguchi-Kuramoto model: numerical simulations in a finite SF and ER network}
\label{SK in finite network}
The Sakaguchi-Kuramoto model \cite{Sakaguchi:1986} on complex networks considered for the present study consists of $N$ coupled oscillators whose phases $\theta_i(t) (i = 1\dots N)$ are driven by the dynamic equations
 \begin{eqnarray}\label{eqn1}
\frac{d\theta_i}{dt} &=& \omega_i +\lambda\sum_{j=1}^{N} A_{ij}\sin(\theta_j - \theta_i-\alpha), ~i = 1\dots N,
\end{eqnarray}
where $\omega_i$ represents the natural frequency of the $i^{th}$ oscillator, $A_{ij}$ is the $ij^{th}$ element of the adjacency matrix $A = (A_{ij})_{N\times N}$ such that $A_{ij} = 1$ if $i^{th}$ and $j^{th}$ oscillators are connected and $A_{ij} = 0$ otherwise, $\alpha$ is the phase-lag parameter whose value lies in the range $0\leq \alpha < \frac{\pi}{2}$ and $\lambda$ is the coupling strength. In the limit $\alpha \rightarrow 0$, the model described by (\ref{eqn1}) converges to the classical Kuramoto model~\cite{Kuramoto}. In most of the studies, $\omega_i$'s are taken randomly from some unimodal distribution $g(\omega)$. However, in this paper, we assume that degrees ($k_i, i = 1\dots N$) and natural frequencies  ($\omega_i, i = 1\dots N$) of the oscillators are linearly related. Hence, without loss of generality we take $\omega_i = k_i~(i = 1\dots N)$, where $k_i$ is the degree of the $i^{th}$ node i.e. $g(\omega) = P(k)$, the degree distribution of the network~\cite{jesus:prl106_2011, Boccaletti:physrep2016,Leyva:screport2013,Peron:pre86_2012,Pinto:2015,Coutinho:pre87_2013}. 
We now numerically simulate the SK model both on SF and ER networks with $N = 1000$ for different values of $\alpha$. For numerical simulation, system~(\ref{eqn1}) is integrated using fourth order Runge-Kutta scheme with time step $\delta t = 0.001$. To quantify the level of synchronization in the system we use the equation
\begin{eqnarray}
r(t)e^{i\psi(t)} = \frac{\sum_{j=1}^{N} k_{j}e^{i\theta_j}}{\sum_{j=1}^{N} k_{j}},
\label{op}
\end{eqnarray}
where $r(t)$ and $\psi(t)$  respectively denote the order parameter and the average phase of the collective dynamics at time $t$. The values of the order parameter $r(t)$ varies in the range $0\leq r(t)\leq 1$. The order parameter takes the value $r(t) = 0$ for incoherent solution, while $r(t) = 1$ indicates fully synchronized state of the system. 
Figure \ref{simu:sf_n1000} shows the effect of phase frustration parameter $\alpha$ on transition to synchronization measured by the time averaged order parameter $r$ for forward and backward continuations as a function of coupling strength $\lambda$ in a scale-free network of size $N = 1000$, $\gamma = 2.8$ and mean degree $\langle k \rangle \sim30$. For $\alpha = 0$, a first-order transition or ES with a large hysteresis loop is observed as reported in \cite{jesus:prl106_2011}. The width of the hysteresis loop is decreased with the increase of the value of $\alpha$ (see Fig.\ \ref{simu:sf_n1000}(b) and (c)). As the value of $\alpha$ crosses a threshold, ES is annihilated  and a second order transition is observed (see Fig.\ \ref{simu:sf_n1000}(d)). We also observe as the value of $\alpha$ increases, critical coupling strength ($\lambda_c$) for the onset of synchronization decreases gradually i.e $\lambda_c(\alpha=0)>\lambda_c(\alpha=0.1)>\lambda_c(\alpha=0.2)>\lambda_c(\alpha=0.5)$. So for SF network, the width of explosive synchronization regime is decreased and the onset of synchronization is promoted by the phase-frustration parameter ($\alpha$). On the other hand we do not observe first order phase transition in ER network $(N = 1000)$  and the onset of synchronization is slightly pushed towards higher value of the coupling strength (see Fig.\ \ref{simu:er_n1000})  as if phase-lag delays the onset of synchronization  in this network i.e $\lambda_c(\alpha=0)<\lambda_c(\alpha=0.8)<\lambda_c(\alpha=1.2)$. Therefore,  the parameter $\alpha$ along with suppressing the ES is enhancing the onset of synchronization in SF network and on the other hand, it is inhibiting the same for ER networks. This is rather surprising and interesting observation, hence demands detailed investigation. In the next section, we investigate the onset of synchronization in detail by performing mean-field analysis of the system using annealed network approximation.

\section{Mean-field analysis}
Following mean-field approach proposed in \cite{ichinomia:pre_2004}, let the density of the nodes with phase $\theta$ at time $t$ for a given degree $k$ be given by the function $\rho(k,\theta,t)$, and it is normalized as  
\begin{eqnarray}
\int_{0}^{2\pi} \rho(k,\theta,t)d\theta =1.\label{density}
\end{eqnarray}
We assume that there is no degree correlation between the nodes of the network and therefore the probability that a randomly chosen edge is attached to a node with degree $k$ and phase $\theta$ at time $t$ can be written as 
\begin{equation}
\frac{kP(k)\rho(k,\theta,t)}{\int kP(k)dk}.
\end{equation}
In the continuum limit, Eq.\ \ref{eqn1}  can be written as 
\begin{eqnarray}
\frac{d\theta(t)}{dt} & = & \omega  + \frac{\lambda k}{\langle k \rangle} \int dk' \int d\theta' k' P(k') \rho(k',\theta',t) \times \nonumber\\
&& \sin(\theta' -\theta -\alpha),
\label{eqn3}
\end{eqnarray}
where $\langle k\rangle = \int kP(k)dk$ is the mean degree of the network. 
Now for the conservation of the oscillators for Eq.\ (\ref{eqn1}), the density function $\rho$ satisfies the continuity equation
\begin{equation}
\frac{\partial \rho}{\partial t} + \frac{\partial}{\partial \theta}(\rho v) = 0,\label{continuity}
\end{equation}
where $v$ is the right hand side of the Eq.\ (\ref{eqn3}). 

To measure the macroscopic behavior of the oscillators, in the thermodynamic limit ($N \rightarrow \infty$) we consider the order parameter $r$ given by \cite{ichinomia:pre_2004} 
\begin{eqnarray}
r e^{i\psi} & = & \frac{1}{\langle k\rangle} \int dk \int d\theta k P(k)\rho(k,\theta,t) e^{i\theta},\label{eqn4}
\end{eqnarray}
where $\psi$ is the average phase of the oscillators and the value of $r$ varies in the range $0\le r \le 1$. Therefore, using (\ref{eqn4}), we rewrite the Eq.\ (\ref{eqn3}) as 
 \begin{eqnarray}
\frac{d\theta}{dt} & = & 
\omega + \lambda k r \sin(\psi -\theta -\alpha).
\label{eqn5}
\end{eqnarray}
In the present study we consider $\omega_i = k_i (i = 1\dots N)$. To derive the self-consistent equation we set the global phase $\psi(t) = \Omega t$ where $\Omega$ is the group angular velocity and introduce a new variable $\phi$ with $\phi(t)=\theta(t)-\psi(t)+\alpha$. In terms of this new variable, equation (\ref{eqn5}) can be written as 
\begin{eqnarray}
\frac{d\phi}{dt} & = & 
k- \Omega - \lambda k r \sin(\phi).\label{eqn6}
\end{eqnarray}
Then the equation of continuity (\ref{continuity}) takes the form
 \begin{eqnarray}
 \frac{\partial}{\partial t} \rho(k,\phi,t) + \frac{\partial}{\partial \phi} [v_\phi \rho(k,\phi,t)]=0, \label{eqn_cont}
 \end{eqnarray}
where $v_\phi = \frac{d\phi}{dt}$. In the steady state, we have 
$\frac{\partial}{\partial t} \rho(k,\phi,t) =0$. 
\noindent Therefore, steady state solution for the density function $\rho$ is given by
\begin{eqnarray}
 \rho(k,\phi)=\begin{cases}
\delta \left(\phi-arc\sin{\left(\frac{k-\Omega}{k\lambda r}\right)}\right), & \abs*{\frac{k-\Omega}{k\lambda r}} \leq 1 \\
\frac{A(k)}{k-\Omega -k\lambda r \sin(\phi)}, & \abs*{\frac{k-\Omega}{k\lambda r}} > 1,
\end{cases}
\end{eqnarray} 
 where $\delta$ is the Dirac delta function and $A(k)$ is the normalization constant given by $A(k) = \frac{\sqrt{(k-\Omega)^2 - (\lambda r k)^2}}{2\pi}$.
The first solution corresponds to the synchronous state and second solution is due to desynchronous state. Hence the order parameter can be rewritten as
\noindent 
 \begin{eqnarray}
r &=& \frac{1}{\langle k\rangle} \int \bigg[\int_{k_{min}}^{\infty}  dk k P(k) \rho(k,\phi) e^{i(\phi -\alpha)}\times \nonumber\\
 && H\bigg(1-\abs*{\frac{k-\Omega}{k\lambda r}}\bigg)\nonumber\\
&& + \int_{k_{min}}^{\infty} dk k P(k) 
 \rho(k,\phi) e^{i(\phi-\alpha)}\times\nonumber\\
 &&  H\bigg(\abs*{\frac{k-\Omega}{k\lambda r}}-1\bigg) \bigg] d\phi, 
 \label{r_d_l}
 \end{eqnarray}
\noindent where $H$ denotes heaviside function. Here the first part of right hand side of Eq.\ (\ref{r_d_l}) gives the contribution of locked oscillators  and the second part denotes the contribution of drift oscillators  to the order parameter $r$.

Therefore, the contribution of locked oscillators to the order parameter is
\begin{eqnarray}
r_{1}&=&\bigg[\frac{\cos \alpha}{\langle k\rangle}\int_{k_{min}}^{\infty} k P(k) \sqrt{1-\left(\frac{k-\Omega}{\lambda r k}\right)^2} dk + \frac{\sin \alpha}{\langle k\rangle} \times\nonumber\\
&&  \int_{k_{min}}^{\infty} k P(k) \frac{k-\Omega}{\lambda r k} dk \bigg]H\left(1-\abs*{\frac{k-\Omega}{k\lambda r}}\right)  \nonumber\\
&&-i\bigg[ \frac{\sin \alpha}{\langle k\rangle}\int_{k_{min}}^{\infty} k P(k) \sqrt{1-\left(\frac{k-\Omega}{\lambda r k}\right)^2} dk -\frac{\cos \alpha}{\langle k\rangle} \times \nonumber \\
&&\int_{k_{min}}^{\infty} k P(k) \frac{k-\Omega}{\lambda r k}dk \bigg]H\left(1-\abs*{\frac{k-\Omega}{k\lambda r}}\right),
\label{r_lock}
 \end{eqnarray}
 and that of the drift oscillators is 
 \begin{eqnarray}
r_{2}&=&\frac{(\sin \alpha +i\cos \alpha)}{\langle k\rangle}\int_{k_{min}}^{\infty} dk \frac{k-\Omega}{\lambda r} P(k) \times \nonumber\\
&& \left[1-\sqrt{1-\left(\frac{\lambda r k}{k-\Omega}\right)^2}  \right] H\left(\abs*{\frac{k-\Omega}{k\lambda r}}-1\right).~~~~
\label{r_drift}
 \end{eqnarray}
 Hence we get $r=r_1 +r_2$, where $r_1$ and $r_2$ are given by Eq.\ (\ref{r_lock}) and Eq.\ (\ref{r_drift}) respectively. 
 Comparing the real and imaginary parts we get
 \begin{eqnarray}
r\langle k\rangle&=&\cos \alpha\int_{k_{min}}^{\infty} dk k P(k) \sqrt{1-\left(\frac{k-\Omega}{\lambda r k}\right)^2} \times  \nonumber\\
&&H\left(1-\abs*{\frac{k-\Omega}{k\lambda r}}\right) + \frac{\sin \alpha}{\lambda r}(\langle k\rangle- \Omega) \nonumber \\
&&-\sin \alpha \int_{k_{min}}^{\infty} dk \frac{k-\Omega}{\lambda rk} k P(k) \times   \nonumber\\  
&& \sqrt{1-\left(\frac{\lambda r k}{k-\Omega}\right)^2}  H\left(\abs*{\frac{k-\Omega}{k\lambda r}}-1\right),  
\label{real_r}
 \end{eqnarray}
and
 \begin{eqnarray}
\langle k\rangle-\Omega&=&\int_{k_{min}}^{\infty} dk (k-\Omega)P(k) \sqrt{1-\left(\frac{\lambda r k}{k-\Omega}\right)^2}  \times \nonumber\\
&& H \left(\abs*{\frac{k-\Omega}{k\lambda r}}-1\right) +\nonumber \\
&&\lambda r \tan \alpha \int_{k_{min}}^{\infty} dk  k P(k) \sqrt{1-\left(\frac{k-\Omega}{\lambda r k}\right)^2} \times \nonumber\\
&& H\left(1-\abs*{\frac{k-\Omega}{k\lambda r}}\right) . 
\label{im_r}
 \end{eqnarray}
 
To simplify the calculations we introduce a variable $x=\lambda r$ and substituting in Eq.\ (\ref{real_r} and Eq.\ (\ref{im_r}) we obtain the following two equations 
 \begin{eqnarray}
\langle k\rangle -\Omega(x)& =&\int_{k_{min}}^{\infty}  (k-\Omega(x))P(k) \sqrt{1-\left(\frac{x k}{k-\Omega(x)}\right)^2}  \nonumber\\ 
&&\times H\left(\abs*{\frac{k-\Omega}{xk}}-1\right)dk+x \tan \alpha \times  \nonumber \\
&&  \int_{k_{min}}^{\infty} dk k P(k) \sqrt{1-\left(\frac{k-\Omega(x)}{x k}\right)^2}  \nonumber\\
&& \times H\left(1-\abs*{\frac{k-\Omega}{x k}}\right)
 \label{omeg_x}
 \end{eqnarray}
and
 \begin{eqnarray}
 \mathrm{R}(x)=\frac{x}{\lambda}. 
 \label{r_x_lambda}
 \end{eqnarray}
The function $\mathrm{R}(x)$ is given by
 \begin{eqnarray}
\mathrm{R}(x)\langle k\rangle&=&\cos \alpha\int_{k_{min}}^{\infty} dk k P(k) \sqrt{1-\left(\frac{k-\Omega}{x k}\right)^2} \times \nonumber\\ 
&& H\left(1-\abs*{\frac{k-\Omega}{x k}}\right)+ \frac{\sin \alpha}{x}(\langle k\rangle- \Omega) -\sin \alpha \nonumber \\
&& \times  \int_{k_{min}}^{\infty} dk \frac{k-\Omega}{x k} k P(k)\sqrt{1-\left(\frac{x k}{k-\Omega}\right)^2} \times \nonumber\\
&& H\left(\abs*{\frac{k-\Omega}{x}}-1\right).
\label{r_x} 
\end{eqnarray}
The equations~(\ref{omeg_x})-~(\ref{r_x}) are self-consistent. For a given network, $\Omega(x)$ can be calculated for different values of $x$ using equation~(\ref{omeg_x}) which can be used to compute the value of the function $\mathrm{R}(x)$ from equation~(\ref{r_x}). Then $r$ can be calculated for a given value of $\lambda$ from the intersection of the graphs of the functions $\mathrm{R}(x)$ and $\frac{x}{\lambda}$. Note that in the limit $\alpha\rightarrow 0$, the self consistent equations~(\ref{omeg_x})-~(\ref{r_x}) naturally converge to the similar equations as derived in~\cite{Coutinho:pre87_2013} for Kurmaoto model. We can now conveniently use these three equations to analyze the transition to synchronization in complex networks in detail. In the following section we consider two SF and one ER networks to demonstrate this.

\section{Analytical and Numerical results: A comparison}
\begin{figure}
   \includegraphics[height=!,width=0.48\textwidth]{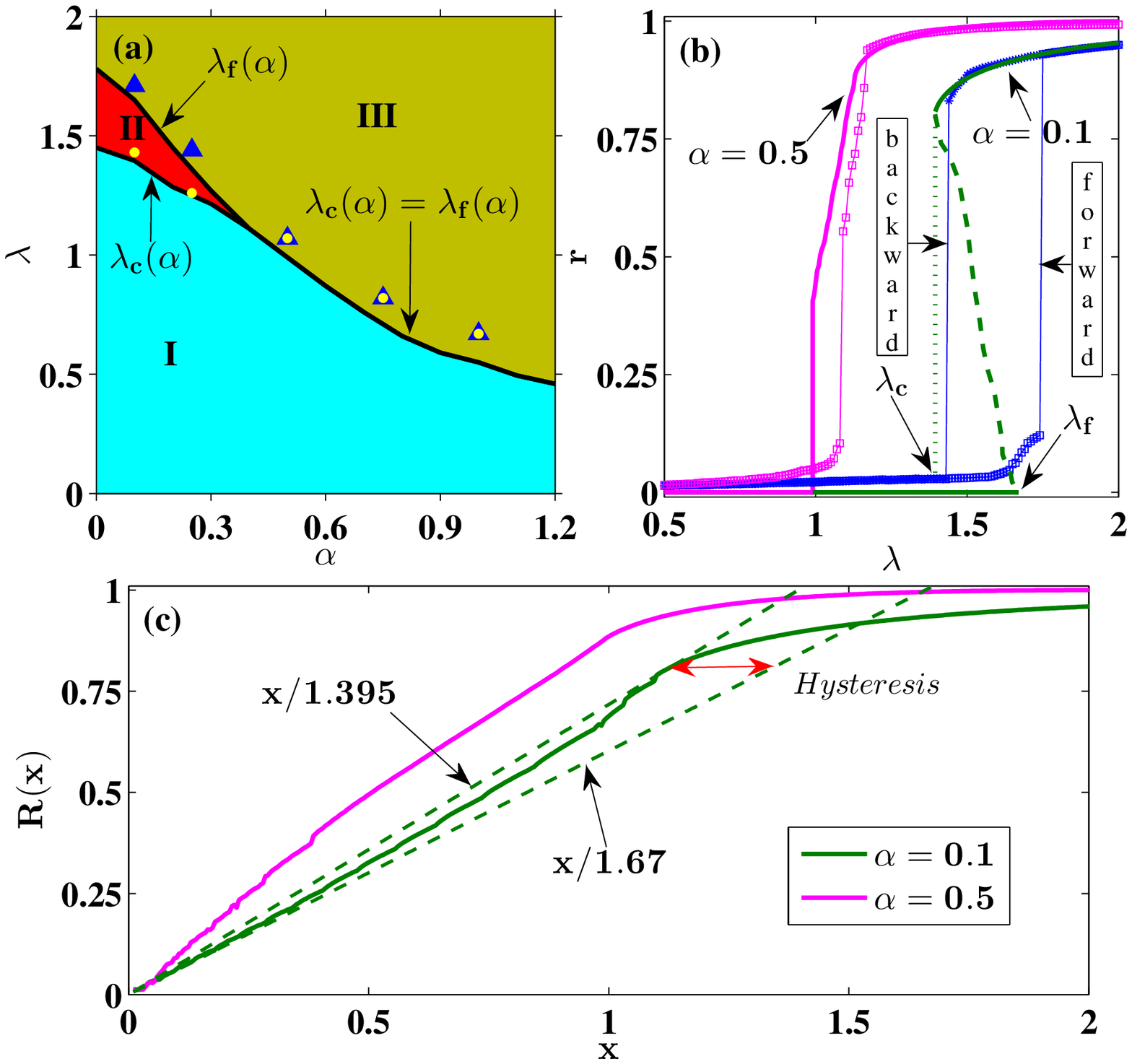}
   \caption{Effect of $\alpha$ on transition to synchronization for a scale-free network of size $N = 10000$ with exponent $\gamma = 2.7$ computed using the equations~(\ref{omeg_x})-(\ref{r_x}). (a) Phase diagram on $\alpha - \lambda$ plane shows  asynchronous (cyan, $r\sim 0$), partially synchronous (green) and  hysteresis (red) regions delimited by the solid back curve marking the critical coupling strength for spontaneous transition to synchrony (asynchrony) during forward ($\lambda_f$) and backward ($\lambda_c$) continuation respectively.  Note that ($\lambda_f = \lambda_c$) when the value of $\alpha$ is greater than a threshold $(\alpha > 0.37)$. Blue triangles and yellow dots are calculated through numerical simulation during forward and backward continuation respectively. Blue triangles and yellow dots are marged to eachother beyond the threshold of alpha. (b) Order parameter $r$ as a function of the coupling strength $\lambda$ for two values of $\alpha$. Solid magenta (for $\alpha = 0.5$) and green (for $\alpha = 0.1$) curves represent analytically calculated order parameters corresponding stable states and dashed green curve represents that corresponding to metastable state indicating the width of ES. Square marked solid curves are obtained from numerical integration (magenta for $\alpha=0.5$ and blue for $\alpha=0.1$). (c) Graph of $\mathrm{R(x)}$ as a funtion of $x$ for $\alpha = 0.1$ (solid green curve) and $0.5$ (solid magenta curve). The upper and lower dashed green lines  correspond to $\lambda_c(0.1)$ and $\lambda_f(0.1)$  indicating the width of the hysteresis (shown by red arrow).} \label{Fig:gamma2p7}
 \end{figure}

To investigate transition to synchrony in SK model on complex networks in detail, first we  consider an uncorrelated  SF network of size  $N=10000$, degree distribution exponent $\gamma = 2.7$ and  mean degree  $\langle k \rangle \sim 30$. The  system follows a degree-frequency correlation i.e. $\omega_i = k_i~(i = 1\dots N)$.   Now using  the self-consistent equations~(\ref{omeg_x})-(\ref{r_x}), the $\lambda-\alpha$  space  is divided into three qualitatively different regions (see figure~\ref{Fig:gamma2p7}(a)). Region I (cyan) represents the desynchronized regime ($r \sim 0$), in the region II (red) the system shows hysteresis with one metastable state and two stable states while synchronization appears in region III (green) spontaneously with $r > 0$. The boundaries (solid black curves) delimiting different regions are determined from the intersection points of the functions $\mathrm{R}(x)$ and $\frac{x}{\lambda}$ for  $0\leq \lambda \leq 2$ at each value of $\alpha$. For each lower values of $\alpha$ ($0\leq\alpha < 0.37$, corresponding to the red region in figure~\ref{Fig:gamma2p7}(a)) as the value of $\lambda$ varied slowly from  $0$ to $2$, the number of intersections between the graphs of $\mathrm{R}(x)$ and $\frac{x}{\lambda}$ first jumps from $1$ to $3$ and then $3$ to $2$. 
The value of $\lambda$ corresponding to the first jump in the number of intersections between the graphs is denoted by $\lambda_c(\alpha)$ and that for second jump by $\lambda_f(\alpha)$.  Therefore, in the range $\lambda_c(\alpha)< \lambda < \lambda_f(\alpha)$ there are three points of intersection which include the trivial solution $x= 0$ and two other nontrivial solutions.  As argued in~\cite{Coutinho:pre87_2013}, in this range, the trivial solution $x = 0$ and the largest $x$ correspond to stable and the intermediate $x$ corresponds to metastable state.  In this region marked by red color in the figure~\ref{Fig:gamma2p7}(a) first order transition takes place. On the other hand, for $\alpha > 0.37$ only one jump from $3$ to $2$ in the number of intersections between the curves $R(x)$ and $\frac{x}{\lambda}$ are observed and in that case we denote the corresponding value of $\lambda$ by $\lambda_c(\alpha)$. The trivial solution $x = 0$ looses stability at the boundary $\lambda_c(\alpha)$ in this case and the transition is second order. In the entire range of $\alpha$ considered in figure~\ref{Fig:gamma2p7}(a), $\lambda_c(\alpha)$ gives the lower boundary of transition to synchrony during backward continuation while $\lambda_f(\alpha)$ gives the same for forward continuation. Note that for $\alpha > 0.37$ both are same and denoted only by $\lambda_c(\alpha)$. Therefore, the solid black curves denoted by $\lambda_f(\alpha)$ and $\lambda_c(\alpha)$ in the figure~\ref{Fig:gamma2p7}(a) give the critical couplings that demarcate the transition from $r\sim 0$ to $r \rightarrow 0^+$ regimes during forward and backward continuations respectively. For more clarity, we show the graphs of the function $\mathrm{R}(x)$ in figure~\ref{Fig:gamma2p7}(c) for $\alpha = 0.1$ (solid green curve) and $0.5$ (solid magenta curve).  The value $\alpha = 0.1$ corresponds to first order transition regime while $\alpha = 0.5$ is taken from the second order transition regime. From the curvature of the solid magenta curve it can be easily understood that it will intersect the straight lines $\frac{x}{\lambda}$ either at one point (for low values of $x$) and at two points for higher values of $x$. The intersection points in turn give the value of $r$ for each value of $\lambda$.  The graph of $r$ as a function of $\lambda$ is shown in figure~\ref{Fig:gamma2p7}(b) with solid magenta curve and second order nature of the transition is clear from it.  On the other hand, the curvature of the solid green curve drawn for $\alpha = 0.1$ in figure~\ref{Fig:gamma2p7}(c) allows three points of intersection with the graphs of   
$\frac{x}{\lambda}$ drawn between the dashed green curves shown in the figure~\ref{Fig:gamma2p7}(c). So $\lambda_f(0.1) = 1.67$ and $\lambda_c(0.1) = 1.395$ in this case. Corresponding values of $r$ have been shown in figure~\ref{Fig:gamma2p7}(b) with green curves. Solid green curves stand for the stable solutions while dashed green curve stands for metastable state. 
\begin{figure}
 \includegraphics[height=!,width=0.50\textwidth]{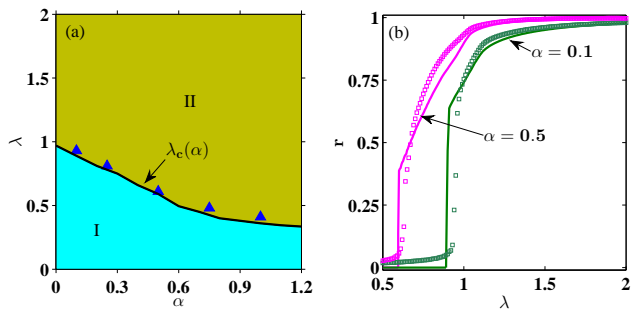}
  \caption{Effect of $\alpha$ on transition to synchronization for a scale-free network of size $N = 10000$ with exponent $\gamma = 3.2$ computed using equations~(\ref{omeg_x})-(\ref{r_x}). (a) Phase diagram on $\alpha - \lambda$ plane shows  asynchronous (cyan, $r\sim 0$) and partially synchronous (green, $r > 0$) regions delimited by the solid back curve marking the critical coupling strength ($\lambda_c$) for spontaneous transition to synchrony. Blue triangles represent the critical coupling strength obtained from numerical simulations. (b) Variation of the order parameter $r$ with coupling strength $\lambda$ for two values of $\alpha$. Solid curves (green and magenta) are drawn based on analytical approach and squares (green and magenta) are obtained from numerical simulations.} \label{Fig:gamma3p2}
\end{figure} 
 
To validate the analytical results,  we take the same degree-frequency correlated SF network ($N = 10000$, $\gamma = 2.7$ and $\langle k \rangle \sim 30$)  and numerically integrate SK model both forward and backward directions in the range $0\leq \lambda \leq 2$ for $\alpha=0.1,~0.25,~0.5,~0.75~\mathrm{and}~1.$    Numerically obtained transition points for forward and backward continuations are shown with filled blue triangles and yellow dots in figure~\ref{Fig:gamma2p7}(a).  Numerically calculated order parameters for $\alpha = 0.5$ and $0.1$ are also plotted in figure~\ref{Fig:gamma2p7}(b) with solid magenta curves marked by squares and solid blue curves marked by square respectively. The results of the numerical simulations are found to match with that of the analytical ones quite nicely and the qualitative features of the transition to synchronization are closely captured by the analytical calculations.

Next we consider another SF network  of same size and average degree ($N=10000$, $\langle k \rangle \sim 30$) but with higher degree-distribution exponent ($\gamma=3.2$).  
Interestingly,  SK model on this network  shows a second order phase transition for all values of $\alpha$. Figure~\ref{Fig:gamma3p2}(a) shows two different regions computed from the equations~(\ref{omeg_x})-(\ref{r_x}) in the $\alpha - \lambda$ space. The cyan and green regions respectively represent de-synchronized ($r \sim 0$) and partially phase locked solutions ($r >0$) separated by a solid black curve which shows critical coupling strength ($\lambda_c(\alpha)$) for the onset of synchronization i.e. a transition from incoherent ($r\sim 0$) to coherent regions ($r > 0$). No hysteresis is observed for this network and transition is purely second order. The transition is also determined by performing numerical simulation of degree-frequency correlated SK model on this SF network and the critical coupling strengths obtained from the simulation for different values of $\alpha$ have been plotted with filled blue triangles in figure~\ref{Fig:gamma3p2}(a) and we find a very close match between analytical and numerical boundaries. Figure~\ref{Fig:gamma3p2}(b) shows the order parameters computed analytically as well as numerically for $\alpha=0.1~\mathrm{and}~0.5$. Analytically calculated order parameters are shown by the solid curves (solid green curve for $\alpha=0.1$ and magenta curve $\alpha=0.5$) while the numerically calculated order parameters are shown with green and magenta squares respectively. For this network also the match between the analytical and numerical results is quite close.
\begin{figure}
  \includegraphics[height=!,width=0.50\textwidth]{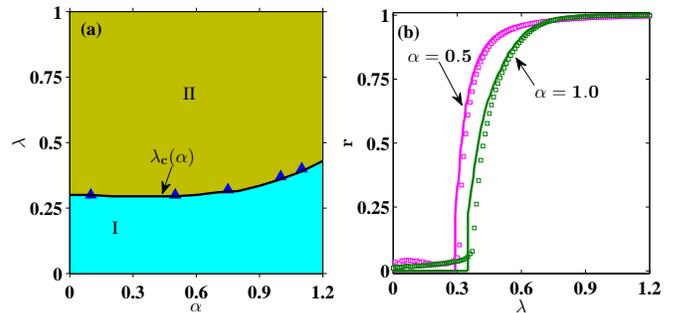}
  \caption{Effect of $\alpha$ on transition to synchronization for an ER of size $N = 10000$ computed using equations~(\ref{omeg_x})-(\ref{r_x}). (a) Phase diagram on $\alpha - \lambda$ plane shows  asynchronous (cyan, $r\sim 0$) and partially synchronous (green, $r > 0$) regions delimited by the solid back curve marking the critical coupling strength ($\lambda_c$) for spontaneous transition to synchrony. Blue triangles represent the critical coupling strength obtained from numerical simulations. (b) Variation of the order parameter $r$ with coupling strength $\lambda$ for two values of $\alpha$. Solid curves (green and magenta) are drawn based on analytical approach and squares (green and magenta) are obtained from numerical simulations.} \label{Fig:er}
\end{figure} 
    
We further explore the impact of phase-frustration parameter $\alpha$ on transition to synchrony in  SK model on ER network. We consider here an ER network of size $N = 10000$ and mean degree $\langle k \rangle \sim 30$. Once again we compute a phase diagram on $\alpha-\lambda$ plane (see figure~\ref{Fig:er}(a)) using the self-consistent equations~(\ref{omeg_x})-(\ref{r_x}). Here the system undergoes second order phase transition for all values of $\alpha$. 
The solid black curve separates the de-synchronization region (I in cyan ) from synchronization regime (II in green) revealing the behavior of the critical coupling. The filled blue triangles shown in the figure~\ref{Fig:er}(a) represent the critical coupling strength computed numerically. Figure~\ref{Fig:er}(b) shows the order parameters calculated analytically as well as through numerical simulations for two values of the phase frustration parameter $\alpha$. So for this ER network the analytically and numerically obtained results match satisfactorily.
  
From the above investigation we observe that degree-frequency correlated SK model displays diverse transition behavior depending on the phase frustration parameter and network structure. 
The system exhibits first order transition or explosive synchronization when the network is SF ($\gamma<3$) and the value of $\alpha$ is small. As the value of $\alpha$ is increased, the width of the hysteresis loop is decreased and eventually hysteresis loop is annihilated at a critical  value $(\alpha_c)$ of $\alpha$. The system displays only second order transition to synchrony for $\alpha > \alpha_c$. The system also shows second order phase transition for SF networks with $\gamma > 3$ and ER networks. We would like to highlight another crucial point here which we have already mentioned in Section:\ref{SK in finite network}. The value of the critical coupling strength $\lambda_c(\alpha)$ is decreased for  SF network ($\gamma\le3$ or $\gamma>3$) when the value of the frustration parameter ($\alpha$) is increased (see figure~\ref{Fig:gamma2p7}(a) and figure~\ref{Fig:gamma3p2}(a)) whereas in ER network the critical coupling is increased slowly if we increase the value of $\alpha$ (see figure~\ref{Fig:er}(a)) i.e. $\alpha$ promotes the onset of  synchronization in SF networks while inhibits the same for ER networks. 
\begin{figure}[h]
\includegraphics[height=!,width=0.50\textwidth]{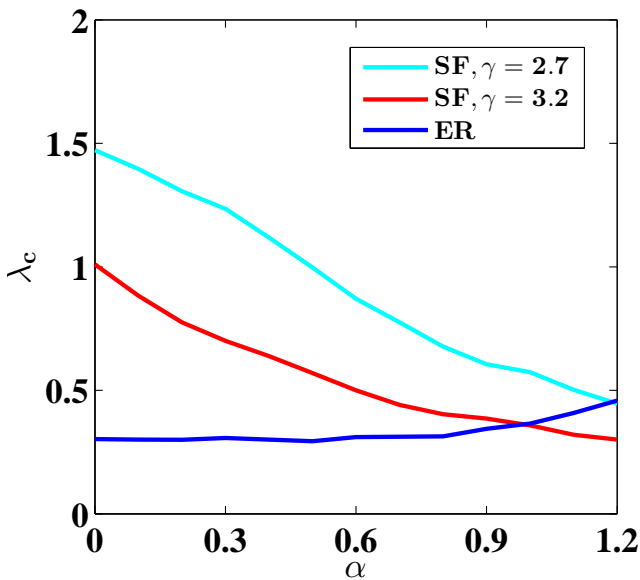}
\caption{Critical coupling strength as a function of $\alpha$ for two SF and one ER networks of size $N = 10000$ and mean degree $\langle k\rangle = 30$ computed from equations~(\ref{Eq:cric_lambda}) and~(\ref{Eq:cric_omega}). Solid cyan and red curves are used for SF networks with $\gamma = 2.7$ and $3.2$ respectively while solid blue curve used for ER network.} \label{Fig:alpha_on_SF_ER}
\end{figure} 

To understand such interesting behavior, we analytically derive the critical coupling strength $\lambda_c(\alpha)$ by considering $r \rightarrow 0^+$  in the equations~(\ref{omeg_x}) and~(\ref{r_x_lambda}) (detailed calculation is shown in the section \ref{Appendix}). The expression for $\lambda_c (\alpha) $  is given by
\begin{eqnarray}
\lambda_c(\alpha) = \frac{2\langle k\rangle \cos \alpha}{\pi \Omega_c^2 P(\Omega_c)},
\label{Eq:cric_lambda}
\end{eqnarray}
where the critical frequency $\Omega_c$ can be determined from the equation
\begin{eqnarray}
\pi\Omega_c^2 P(\Omega_c)\tan \alpha=\int_{k_{min}}^{\infty} \frac{k^2 P(k)}{k-\Omega_c} dk.
\label{Eq:cric_omega}
\end{eqnarray}
From equation~(\ref{Eq:cric_lambda}) we observe that $\lambda_c(\alpha)$ implicitly depends on both structural as well as dynamic properties of the networks in a complicated way. We now calculate $\lambda_c(\alpha)$ as a function of $\alpha$ using the equations~(\ref{Eq:cric_lambda}) and (\ref{Eq:cric_omega}). Figure~\ref{Fig:alpha_on_SF_ER} shows the variation of critical coupling strength $\lambda_c(\alpha)$ for two SF networks and one ER network used in the this section. The graphs of $\lambda_c(\alpha)$ for three networks (see figure~\ref{Fig:alpha_on_SF_ER}) computed from the equations (\ref{Eq:cric_lambda}) and (\ref{Eq:cric_omega}) are similar to the ones obtained from the equations~(\ref{omeg_x})-(\ref{r_x}) (see figures~\ref{Fig:gamma2p7}(a),~\ref{Fig:gamma3p2}(a) and~\ref{Fig:er}(a)).

\section{Conclusions}
We have performed analytical as well as numerical investigations to understand the synchrony behaviour in degree-frequency linearly related Sakaguchi-Kuramoto model on complex networks using mean-field analysis. Analytically we have derived self-consistent equations involving group angular velocity and order parameter of the system which successfully explain different types of transition to synchronization in the presence of phase-frustration parameter. For demonstration, we consider degree-frequency correlated SK model on uncorrelated SF and ER networks. From the analytical approach we find first order transition to synchronization or ES in SK model on SF networks with $\gamma < 3$ for low values of the frustration parameter $\alpha$  while for $\gamma > 3$ we find only second order transition to synchronization for all values of $\alpha$. However, in SF networks, although the phase-frustration parameter $\alpha$ annihilates ES and promotes the transition to synchronization i.e decreases the critical coupling strength. On the other hand, for ER networks, $\alpha$ inhibits transition to synchrony. Using this analytical approach, we determine the order parameters both for SF and ER networks. The critical coupling strength for the onset of synchronization for forward and backward continuation in both the networks have also been determined. We also perform a detailed numerical simulation to validate the analytical results. The numerical results show close agreement with that of the analytical ones for networks of large size. 
\section{Acknowledgements}
\par The authors would like to thank Syamal Dana for interesting comments and suggestions. PK acknowledges support from  DST, India under the DST-INSPIRE scheme (Code: IF140880). CH is supported by the CHE/PBC, Israel.

\section{Appendix A: Calculation of critical coupling and critical $\Omega$}
\label{Appendix}
From equation~(\ref{im_r}) using Taylor's series expansion we get
\begin{eqnarray}
\langle k\rangle -\Omega =\lambda r \tan \alpha \int \displaylimits_{\frac{\Omega}{(1+\lambda r)}}^{\frac{\Omega}{(1-\lambda r)}} k P(k) \sqrt{1-\left(\frac{k-\Omega}{\lambda r k}\right)^2}  dk  \nonumber \\ + \int_{k_{min}}^{\infty} (k-\Omega)P(k) \left\lbrace 1-{\frac{(\lambda r k)^2}{2(k-\Omega)^2}}\right\rbrace.
 \end{eqnarray}
Taking the limit $r\rightarrow 0^+$ we can find
 \begin{eqnarray}
 \pi\Omega_c^2 P(\Omega_c)\tan \alpha=\int_{k_{min}}^{\infty} \frac{k^2 P(k)}{k-\Omega_c} dk,
 \label{cric_omegan}
 \end{eqnarray}
where $\Omega_c$ the critical group angular velocity at the onset of synchronization.

Now combining equations~(\ref{real_r}) and~(\ref{im_r}) we get 
 \begin{eqnarray}
r\langle k\rangle=\frac{1}{\cos \alpha}\int_{\frac{\Omega}{1-\lambda r}}^{\frac{\Omega}{1-\lambda r}} k P(k) \sqrt{1-\left(\frac{k-\Omega}{\lambda r k}\right)^2}  dk. 
\label{cric_coup1}
 \end{eqnarray}
Substituting $\frac{k-\Omega}{\lambda r}=y$, the equation~(\ref{cric_coup1})reduces to
\begin{eqnarray}
\langle k\rangle&=&\frac{\lambda}{\cos \alpha}\int_{\frac{-\Omega}{1-\lambda r}}^{\frac{\Omega}{1-\lambda r}} (\Omega+\lambda r y) P(\Omega+\lambda r y)\times\nonumber \\
&& \sqrt{1-\frac{y^2}{(\Omega+\lambda r y)^2}}  dy
\end{eqnarray}
and in the limit $r\rightarrow 0^+$ we get
 \begin{eqnarray}
\lambda_c(\alpha)=\frac{2\langle k\rangle \cos \alpha}{\pi \Omega_c^2 P(\Omega_c)},
\end{eqnarray}
where $\Omega_c$ calculated from equation(\ref{cric_omegan}) and $\lambda_c(\alpha)$ is the critical coupling strength for the onset of synchronization.
\end{document}